\documentclass[a4paper,11pt]{article}
\usepackage[english]{babel} 
\usepackage[T1]{fontenc}
\usepackage{float}
\usepackage{setspace}
\usepackage{appendix}
\usepackage{parcolumns}
\usepackage{stmaryrd}
\usepackage{cite}
\usepackage{times}
\usepackage[OT1]{fontenc}
\usepackage{type1cm}
\usepackage{url}
\usepackage{subfigure}
\usepackage{braket}
\usepackage{graphicx}
\usepackage{sidecap}
\usepackage{xspace}
\usepackage{tikz}
\usepackage{pgflibraryarrows}
\usepackage{pgflibrarysnakes}
\usepackage{indentfirst}
\usepackage[mathscr]{eucal}
\usepackage{geometry}
\usepackage{booktabs}
\usepackage{indentfirst}
\usepackage{bm}
\usepackage{multirow}
\usepackage{dcolumn}
\usepackage{graphicx}
\usepackage{mathrsfs}  
\usepackage{enumerate}
\usepackage[utf8]{inputenc}
\usepackage{amsmath}
\usepackage{amssymb}
\usepackage{amsfonts}
\usepackage{amsthm}
\usepackage{mathrsfs}
\usepackage{makeidx}
\usepackage{graphicx}
\usepackage{ulem}			

\makeatletter
\@addtoreset{equation}{section}

\makeatother

\def\det{{\rm det}}

\newcommand{\be}{\begin{equation}}
\newcommand{\ee}{\end{equation}}
\newcommand{\bea}{\begin{eqnarray}}
\newcommand{\eea}{\end{eqnarray}}
\newcommand{\vs}[1]{\vspace{#1 mm}}

\newcommand{\Lie}{{\cal L}}

\DeclareMathOperator\arctanh{arctanh}

\begin{document}

\vs{8}
\begin{center}
{\Large\bf Wicked metrics}
\vs{6}

{\large
Alessio Baldazzi\footnote{e-mail address: abaldazz@sissa.it}\ 
Roberto Percacci\footnote{e-mail address: percacci@sissa.it}\ 
and Vedran Skrinjar\footnote{e-mail address: vedran.skrin@gmail.com}%
\vs{3}

{ International School for Advanced Studies, via Bonomea 265, I-34136 Trieste, Italy}
{ and INFN, Sezione di Trieste, Italy}

}

\vs{8}
{\bf Abstract}
\end{center}
There are various ways of defining the Wick rotation 
in a gravitational context.
There are good arguments to view it as an analytic continuation 
of the metric, instead of the coordinates.
We focus on one very general definition and argue that it is 
incompatible with the requirement of
preserving the field equations and the symmetries at global level:
in some cases the Euclidean metric cannot be defined on the
original Lorentzian manifold but only on a submanifold.
This phenomenon is related to the existence of horizons,
as illustrated in the cases of the de Sitter and Schwarzschild metrics.

\section{Wick rotation in gravity}

In Quantum Field Theory (QFT) it is customary
to define functional integrals in imaginary (Euclidean) time,
in order to replace oscillatory integrals
by exponentially damped ones, thereby improving
their convergence.
The rotation of the integration contour over energy or time
is called Wick rotation.
In this paper we discuss various issues that arise when one 
considers the Wick rotation for a QFT on a curved spacetime.

In general there will be more than one stationary point for the 
Euclidean action, in which case the functional integral 
should be defined as a sum of Gaussian integrals
around all the regular, finite action solutions 
of the Euclidean field equations, generally known as instantons.
(Note that the instantons will not in general correspond to
solutions of the Lorentzian field equations.)
In the case of quantum gravity, this procedure has been 
developed mainly
by the Cambridge school in the late '70s and early '80s
and is called Euclidean Quantum Gravity \cite{eqg}.

In this approach the Euclidean and Lorentzian spacetimes
are seen as different real sections of a complex manifold.
In practice, one rotates a suitable time coordinate,
as in standard QFT.
This definition of Wick rotation has some well-known shortcomings 
that have been recently summarized in \cite{Visser:2017atf}.
An alternative definition, where the coordinates are kept 
fixed and it is the metric that is analytically continued,
avoids some of these issues.

The main virtue of this alternative definition is that it keeps the
spacetime manifold fixed.
Thus, in the path integral, one would only consider
manifolds that admit a physical, Lorentzian, metric.
Unfortunately, the analytic continuation of the metric
is far from unique.
One may try to fix the ambiguities, or at least to restrict them, 
by imposing some additional desirable properties,
such as mapping local solutions of the Lorentzian field equations
to local solutions of the Euclidean equations,
and preserving the number of Killing vectors.
Our main result is that in some cases these properties
seem to be in conflict with the requirement that the Wick rotation
preserve the manifold.

In the rest of section 1, we will review various definitions
of Wick rotation.
In section 2 we discuss the analytic continuations 
of Minkowski and Anti-de Sitter space.
Sections 3 and 4 deal with the de Sitter and Schwarzschild metrics,
respectively.
Section 5 contains a short discussion, 
where we compare again the continuation of the metric 
with the continuation of the coordinates,
in view of the preceding results.

\subsection{Continuing time}

Assume that spacetime has topology $R\times\Sigma$,
with coordinates $t$ in $R$ and $x^i$ in $\Sigma$.
We further assume that spacetime is static,
namely that there exists a Killing vector that is 
everywhere orthogonal to $\Sigma$.
There exist coordinates where the metric has the form
\be
g_{\mu\nu}(t,x)=\left(
\begin{array}{cc}
g_{00}(x) & 0 \\
0 & g_{ij}(x) 
\end{array}
\right)
\ee
where $g_{00}<0$.
In this case there is a natural time coordinate and
the Wick rotation can be defined in the usual way:
\be
iS_L\Big|_{t\to-it_E}=-S_E\ .
\label{wick}
\ee
For example, for a real scalar field this gives
\be
S_E(\phi)=\frac{1}{2}\int d^dx_E\sqrt{g_E}\,g_E^{\mu\nu}\partial_\mu\phi\partial_\nu\phi
\ee
where
\be
(g_E)_{\mu\nu}=\left(
\begin{array}{cc}
-g_{00} & 0 \\
0 & g_{ij} 
\end{array}
\right)
\ee
is a positive definite metric on an analytically continued manifold.

\noindent
If this definition is extended to more general spacetimes,
several issues arise \cite{Visser:2017atf}.

$\bullet$ {\sl First} we note that time has no physical meaning in GR.
If the Wick rotation is performed on time,
one immediately finds 
that the result depends very strongly on the coordinate system.
Thus for example beginning from the de Sitter metric,
written in three different forms:
the form with flat spatial sections
\be
\label{deSflat}
ds^2=-dt^2+H^{-2}e^{2Ht}\left(dr^2+r^2d\Omega_2^2\right)
\ee
or the form with positively curved spatial sections
\be
\label{deSpos}
ds^2=-d\tau^2+H^{-2}\cosh^2(H\tau)\left(\frac{dr^2}{1-r^2}+r^2d\Omega_2^2\right)
\ee
or the form with negatively curved spatial sections
\be
\label{deSneg}
ds^2=-d\bar\tau^2+H^{-2}\sinh^2(H\bar\tau)\left(\frac{dr^2}{1+r^2}+r^2d\Omega_2^2\right)
\ee
(where $d\Omega_2^2=d\theta^2+\sin^2\theta d\varphi^2$
is the metric of the 2-sphere)
the prescription $t\to -it$ leads to a metric that is either complex,
or positive definite, or again Lorentzian but with opposite signature.

$\bullet$ {\sl Second,} in flat spacetime, the sense of the Wick rotation is fixed by
the requirement that the analytic continuation of the Feynman propagator
of a free particle should not cross the poles in the complex energy plane.
This is related to Feynman's ``$i\epsilon$'' prescription, 
which is a way
to incorporate the notion of causality in the two-point function.
Furthermore, the Euclidean continuation of any correlation functions
must satisfy Osterwalder-Schrader positivity, which is again 
a consequence of causality.  
No such restrictions from causality seem to limit the analytic continuation 
of a time coordinate in a generic Lorentzian manifold.
On the other hand it has been argued that a notion of
causality should be encoded in the functional integral
\cite{Teitelboim:1983fh,causet,Ambjorn:1998xu}.


$\bullet$ {\sl Third,} given that any manifold admits 
a Euclidean metric, 
a definition of the functional integral that started from
Euclidean signature, as advocated in 
Euclidean Quantum Gravity \cite{Gibbons:1976ue}, 
would include a sum over all topologies.
This is a source of various difficulties.
At a fundamental mathematical level,
we are confronted with the fact that even the classification 
of all four-dimensional topologies is impossible \cite{vanmeter}.
This is a major challenge to the definition of a functional integral,
over and above the usual functional analytic issues.
More concretely, numerical simulations of sums over
four-dimensional Euclidean triangulations 
(called ``Euclidean Dynamical Triangulations'' - DT)
have largely failed to produce viable phases 
looking like an extended four-dimensional manifold 
\cite{Bialas:1996wu,deBakker:1996zx,Coumbe:2014nea,Rindlisbacher:2015ewa}.
On the other hand, the existence of a Lorentzian
structure on a given manifold restricts the possible topologies
\cite{hawkingellis}.
Numerical simulations within Causal Dynamical Triangulations (CDT)
have indeed shown that this requirement has a very beneficial 
effect on the functional integral for gravity \cite{Ambjorn:2004qm}.

$\bullet$ {\sl Fourth,} when gravity is viewed as a gauge theory
for the Lorentz group, as in the tetrad formalism,
not only is the signature of the metric changed, but also
the gauge group itself.
This is in sharp contrast to other gauge theories.
This problem is particularly urgent when one couples gravity
to fermions, because the spinor representations are generally different
for different signatures. We will not deal with this issue
in detail here, except for some comments in section 1.6.

\smallskip

For all these reasons we are led to seek a definition 
of Wick rotation that satisfies the following three conditions:
\begin{enumerate}
\item[{\bf(1)}] it does not depend on the coordinates\ ;
\item[{\bf(2)}] causality is taken into account\ ;
\item[{\bf(3)}] the Wick-rotated metric is defined on the same manifold
as the original Lorentzian metric\ .
\end{enumerate}
In the rest of this section we shall discuss two such definitions.
In the rest of the paper we show that even these definitions
are not completely satisfactory.

\subsection{Continuing the lapse}

A better procedure is to analytically continue the
metric instead of time.
One way to do this is to start from an ADM foliation.
Let us assume topology $R\times\Sigma$.
A generic metric and its inverse
can be written locally in the form
\be
g_{(\sigma)\mu\nu}=\left(
\begin{array}{cc}
\sigma N^2+q_{ij}N^i N^j & N_i \\
N_j & q_{ij} 
\end{array}
\right)
\ ,\qquad
g_{(\sigma)}^{\mu\nu}=\left(
\begin{array}{cc}
\frac{1}{\sigma N^2} & -\frac{N^i}{\sigma N^2} \\
-\frac{N^j}{\sigma N^2} & q^{ij}+\frac{N^i N^j}{\sigma N^2}
\end{array}
\right)
\ee
where $q_{ij}$ is a positive definite metric in $\Sigma$.
Note that $\sqrt{g_{(\sigma)}}=\sqrt{\sigma}N\sqrt{q}$.
The metric is Euclidean for $\sigma=1$ and Lorentzian
for $\sigma=-1$.
(One could equivalently take $\sigma=1$ and assume that the sign
of $N^2$ could be negative.)

Now we do not take the absolute value of the determinant
when we construct the action. 
For example, in the scalar case we define
\be
S_{(\sigma)}(\phi)=\frac{1}{2}\int d^dx\sqrt{g_{(\sigma)}}
g_{(\sigma)}^{\mu\nu}\partial_\mu\phi\partial_\nu\phi\ .
\label{scalar}
\ee
Since for Lorentzian metric
($\sigma=-1$) we have $\sqrt{g_{(-1)}}=iN\sqrt{q}$
whereas for Euclidean metric $\sqrt{g_{(1)}}=N\sqrt{q}$,
this action cannot remain real when $\sigma$ changes sign.
For $\sigma=-1$ we define the real Lorentzian action
\be
S_L(\phi)=iS_{(-1)}(\phi)
\label{scal}
\ee
whereas for $\sigma=1$ we define the real Euclidean action
\be
S_E(\phi)=S_{(1)}(\phi)\ .
\label{scae}
\ee
This analytic continuation in $\sigma$ is such that
if we start from $iS_L=-S_{(-1)}$ we end at $-S_{(1)}=-S_E$.
One can similarly check that
\be
S_{(\sigma)}=\frac{1}{4}\int d^dx\sqrt{g_{(\sigma)}}
g_{(\sigma)}^{\mu\nu}g_{(\sigma)}^{\rho\sigma}F_{\mu\rho}F_{\nu\sigma}
\label{maxwell}
\ee
and 
\be
S_{(\sigma)}=\frac{1}{2\kappa^2}\int d^dx\sqrt{g_{(\sigma)}}
(2\Lambda-R(g_{(\sigma)}))
\label{hilbert}
\ee
interpolate between Lorentzian and Euclidean path integrals
for electromagnetism and gravity, always with the identifications
(\ref{scal}) and (\ref{scae}).
In this way we reproduce the result obtained by continuing the time
in the static case, but this procedure is not restricted
to the static case.

\subsection{General procedure}

Every manifold admits a Riemannian (Euclidean) metric
but that there are topological restrictions for the existence of Lorentzian metrics,
namely there must exist a nowhere zero vectorfield \cite{hawkingellis}.
Without loss of generality, such vectorfield can be unit-normalized.
Then, a Lorentzian metric $g_{(L)\mu\nu}$ can be constructed
starting from a Euclidean metric $g_{(E)\mu\nu}$
and a unit vector field $X^\mu$ 
by the formula:
\be
\label{matthew}
g_{(L)\mu\nu}=g_{(E)\mu\nu}-2X_\mu X_\nu\ ,
\ee
where $X_\mu=g_{(E)\mu\nu}X^\nu$.
In the Lorentzian metric, $g_{(L)\mu\nu}X^\mu X^\nu=-1$,
so $X$ is a unit timelike vectorfield.
This formula can be inverted to construct a Euclidean metric out of a given Lorentzian metric and a unit timelike vectorfield.
This can be seen as the result of a continuous deformation \cite{Candelas:1977tt}:
\be
g_{(\sigma)\mu\nu}=g_{(L)\mu\nu}+(1+\sigma) X_\mu X_\nu\ ,
\label{ancont}
\ee
where $\sigma$ varies between $-1$ and $1$.
Clearly $g_{(-1)}=g_{(L)}$ and $g_{(1)}=g_{(E)}$.
For $\sigma=0$ the metric is degenerate:
for any vectorfield $Y^\mu$,
$g_{(0)\mu\nu}X^\mu Y^\nu
=0$.

We note that continuing the lapse is a special case of this
more general procedure,
where $X^\mu$ is the unit normal
to the hypersurfaces of constant time:
$$
X_\mu=(-N,0)\ ;\qquad
X^\mu=\left(\frac{1}{N},-\frac{N^a}{N}\right)\ ,
$$
in the ADM coordinates.
The procedure discussed in this section is more general in that
the vector $X$ is not assumed to be hypersurface-orthogonal.

Let us see how this procedure reproduces the Wick rotation in flat spacetime.
We have $g_{(-1)\mu\nu}=g_{(L)\mu\nu}=\eta_{\mu\nu}$,
so the interpolating metric is
$g_{(\sigma)\mu\nu}=\mathrm{diag}(\sigma,1,1,1)$,
and $g_{(1)\mu\nu}\equiv g_{(E)\mu\nu}=\delta_{\mu\nu}$.
The volume element
$\sqrt{\det g^{(\sigma)}}$ is $\sqrt{\det(\eta_{\mu\nu})}=i$ for $\sigma=-1$ 
and $\sqrt{\det(\delta_{\mu\nu})}=1$ for $\sigma=1$.
With this definition of the Wick rotation, the ``$i$'' in the
exponent in the functional integral comes from taking 
the square root of the determinant of a
metric with Lorentzian signature.
The interpolating actions for scalar, Maxwell and gravitational field
are given again by (\ref{scalar},\ref{maxwell},\ref{hilbert}),
with the identifications (\ref{scal}) for the Lorentzian action
and (\ref{scae}) for the Euclidean action.

\subsection{Complexification}

The two definitions of Wick rotation given in the preceding sections 
give rise to a problem:
if we interpret $\sigma$ as a continuous real parameter
running between $-1$ and $1$, then 
for $\sigma=0$ the metric would become degenerate.
To avoid this, one has to allow $\sigma$ to describe a path in the complex plane.
The question of the contour then arises: does the path pass
above or below the point $\sigma=0$?
To make a choice, we note that the propagator constructed 
with the interpolating metric 
\be
\frac{i}{-g_{(\sigma)\mu\nu}p^\mu p^\nu-m^2}
=\frac{i}{-\sigma E^2-\vec p^2-m^2}
\ee
coincides for $\sigma=-1$ with the causal (Feynman) propagator,
\be
\Delta^F=\frac{i}{E^2-\vec p^2-m^2+i\epsilon}
\ee
if $\sigma$ is given a small negative imaginary part
\be
\sigma=-1-\frac{i\epsilon}{E^2}\ .
\ee
We see that the usual prescription for the choice of integration contour in the
definition of the propagator can be interpreted naturally
as an incipient complexification of the metric.
After allowing $\mathrm{Re}(\sigma)$ to grow from $-1$ to $1$ and letting $\mathrm{Im}(\sigma)$
go back to zero, and taking into account the factor $i$ 
from the volume element, the correlator takes the Euclidean form
\be
\frac{-1}{-g_{(1)\mu\nu}p^\mu p^\nu-m^2}=
\frac{1}{E^2+\vec p^2+m^2}\ .
\ee
The path that we have just described can be deformed into
a path running along the real axis, except for an infinitesimal
semicircle passing above $\sigma=0$.
In the following this path will always be understood.

There is in general no notion of reflection positivity 
in curved spacetime because generically there is no isometry 
that can serve the function of reflection.
At least on static spacetimes, where such a reflection
exists, a suitable generalization of reflection positivity holds \cite{jaffe}.

\subsection{Properties}

The procedures outlined in sects. 1.2-1.3
clearly satisfy the conditions (1)-(2)-(3)
spelled out in section 1.1.
In spite of this, important issues remain,
the most important one being the lack of uniqueness.
The procedure of section 1.2 depends on the choice of a foliation
and the procedure of section 1.3 depends on the choice
of a one-form $X_\mu$:
in both cases there is an infinite dimensional arbitrariness.
One may try to restrict this choice by making additional demands.
For instance, it would be clearly desirable that a definition of 
Wick rotation had the following properties:
\begin{enumerate}
\item[{\bf(4)}] a local solution of the Lorentzian field equations
should map to a local solution of the Euclidean field equations.
For Einstein's equations with a cosmological constant,
this would mean that locally Einstein metrics are mapped to 
locally Einstein metrics.
(The sign of the cosmological constant should be allowed to change.)
\item[{\bf(5)}] if the Lorentzian metric has a Killing vector,
the Euclidean metric should also have a Killing vector.
(In general one would have to allow the algebra of the
Killing vectors to be deformed in the Euclidean continuation, 
as is the case already for flat space.)
\item[{\bf(6)}] a maximally symmetric spacetime should be mapped to a 
maximally symmetric spacetime. 
(Again, one cannot demand the sign of the
curvature to remain the same.)
\end{enumerate}

In point (4) it would be too much to demand that global solutions are
mapped to global solutions.
This is because already for a simple scalar field,
such a property does not hold.
For example on a torus a local solution of the D'Alembert 
equation of the form $e^{ik(x\pm t)}$
maps to a local solution of the Laplace equation of the form
$e^{ikx\mp kt}$, which however does not satisfy the periodicity
conditions. Thus only the constants are global
solutions of the Laplace equation.
All the oscillating solutions do not have a Euclidean analogue.

Let us observe that requirements (4-6)
are automatically satisfied if one interprets the Wick
rotation as a complex change of coordinates.
The main point of this paper will be to show that it is not 
always possible to satisfy all these conditions simultaneously.

\subsection{Other approaches}

We mention here some related ideas that have appeared in
the literature.

Whereas here we view the Wick rotation as a mathematical trick,
one could think of the signature of the metric as being 
dynamically determined.
One can view the metric as an order parameter
whose expectation value breaks the linear group $GL(d)$
to $O(d)$ or $O(d-1,1)$ \cite{Percacci:1990wy,Percacci:2009ij}.
Dynamical mechanisms that determine the signature have been 
discussed in \cite{Carlini:1993up,Dereli:1993pj,Wetterich:2004za,Mukohyama:2013ew,Kehayias:2014uta}.

A definition of Euclidean continuation of spinors
that avoids the doubling issue has been discussed in 
\cite{Mehta:1986mi,Wetterich:2010ni}.

Mathematical results concerning the analytic continuation
of Riemannian manifolds have been discussed in
\cite{Helleland:2015wva,Helleland:2017zks}.


\section{Regular Examples}

\subsection{Minkowski spacetime}

We begin from this rather trivial case, just to show
that it is possible to satisfy all the requirements 1-4,
provided we allow the Killing vectors to be deformed
and their algebra to change (discontinuously)
when one crosses the point $\sigma=0$.
Choosing the one-form $X=dt$ in Minkowski coordinates: 
$$
ds^2(\sigma) = \sigma dt^2 + \sum_{i=1}^{3}(dx^i)^2\ ,
$$
where $-1\leq\sigma\leq1$. 
The Killing vectors are:
\begin{equation}
\footnotesize
\begin{split}
    & P_0 \equiv \frac{1}{\sqrt{|\sigma|}}\partial_0 \hspace{1cm}, \hspace{1 cm} P_i \equiv \partial_i \hspace{3 cm} \mbox{with} 
    \;i=1,2,3\\
    & M_{a b} \equiv x^a \partial_{b} - x^b \partial_{a} \hspace{5 cm} \mbox{with} \;a,b=1,2,3\; \mbox{and}\;a\ne b\\
    & K_a \equiv \frac{1}{\sqrt{|\sigma|}} \left[ x^a \partial_{0} + (1-\sigma)x^0 \partial_{a} \right] \hspace{3 cm} \mbox{with} \;a=1,2,3
\end{split}
\label{genminkill}
\end{equation}
The commutators are:
\begin{equation}
\begin{split}
& \left[ P_i  , P_j \right] = 0 \hspace{3 cm}  
\left[ M_{a b} , P_i \right] =  \delta_{b i} P_a - \delta_{a i} P_b \\
&  \left[ M_{a b} , M_{c d} \right] = -\delta_{a c}M_{b d}+\delta_{a d}M_{b c}+\delta_{c b}M_{a d}-\delta_{b d}M_{a c} \\
& \left[ M_{a b} , K_c \right] = \delta_{b c} K_a - \delta_{a c} K_b \hspace{0.5 cm}  \left[ P_i  , K_a \right] = \delta_{i a} P_0
\\
& \left[ K_a , K_b \right] = \mbox{sign}(-\sigma) M_{a b}  \hspace{1 cm} \left[ P_0  , K_a \right] = \mbox{sign}(-\sigma) P_a 
\end{split}
\label{comgenmin}
\end{equation}
The main point to observe here is that,
as long as $\sigma$ is real, the algebra
remains the same under infinitesimal changes of $\sigma$
but changes discontinuously when $\sigma$ changes sign.
This is because all the metrics with $\sigma<0$
are isometric to the metric with $\sigma=-1$.
The pullbacks of the original Killing vectors by this isometry
are Killing vectors for the deformed metric,
all satisfying the same algebra.
Metrics with $\sigma>0$, however,
are not isometric to the original Minkowski metric
(they are all isometric to the Euclidean metric with $\sigma=1$)
and the boost generators become an additional rotation generator.

\subsection{Anti de Sitter space}

Anti-de Sitter space can be embedded in a flat 5-dimensional
space with metric
$ds^2=-dz_0^2+dz_1^2+dz_2^2+dz_3^2-dz_4^2$.
The embedding equation is
$$
-z^0_2+z^1_2+z^2_2+z^3_2-z^4_2=-r^2\ .
$$
One can choose coordinates $\tau,\chi,\theta,\varphi$, 
defined by
\bea
z_i&=& \,r\,\sinh\chi\,\omega_i \ \ \ \ \ 
\mbox{with}\; i=1,2,3\ \ \mbox{and}\ \ \sum_i\omega_i^2=1 \nonumber\\
z_4&=& \,r\,\cosh\chi\sin\tau,\nonumber\\
z_0&=& \,r\,\cosh\chi\cos\tau\ .
\label{adscoord}
\eea
Aside from the issue of periodicity in the $\tau$ direction,
these coordinates cover the whole manifold.
This embedding gives rise to the metric
\be
\label{AdeSstat}
ds^2=r^2\left(-\cosh^2\chi d\tau^2+d\chi^2+\sinh^2\chi(d\theta^2+\sin^2\theta d\varphi^2)\right)\ .
\ee

The one-form $X=r\cosh\chi d\tau$ has norm $-1$, and can be used in (\ref{matthew})
to generate the Euclidean metric
\be
\label{hypstat}
ds^2=r^2\left(\cosh^2\chi d\tau^2+d\chi^2+\sinh^2\chi(d\theta^2+\sin^2\theta d\varphi^2)\right)\ .
\ee
This is the standard metric on the 4-four-dimensional one-sheeted hyperboloid, 
which is embedded in a five-dimensional Minkowski space with metric
$ds^2=dz_1^2+dz_2^2+dz_3^2+dz_4^2-dz_5^2$
by the condition
$$
z_1^2+z_2^2+z_3^2+z_4^2-z_5^2=-r^2\ .
$$
The coordinates are defined as in (\ref{adscoord}),
except that in the last two lines the trigonometric functions
of $\tau$ are replaced by hyperbolic functions.

The curvature scalar of this space is $R=-12/r^2$, and therefore
it is a solution of Einstein's equations with cosmological constant
$\Lambda=-3/r^2$.
It is maximally symmetric, so the number of Killing vectors
is preserved, but the isometry group changes
from $SO(2,3)$ to $SO(1,4)$.
From this point of view AdS behaves exactly like Minkowski space.

\section{De Sitter space}

De Sitter space in $4$ dimensions has the topology 
of a cylinder $\mathbb{R}\times S^3$.
It can be embedded in a $5$-dimensional Minkowski space
with metric 
$ds^2=-dz_0^2+dz_1^2+dz_2^2+dz_3^2+dz_4^2$ 
by the equation
$$
-z_0^2+z_1^2+z_2^2+z_3^2+z_4^2=H^{-2}\ .
$$
The hyperspherical coordinates 
$\tau$, $\chi$, $\theta$, $\varphi$  
are related to the coordinates of (\ref{deSpos})
by $r=\sin\chi$.
They are related to the embedding coordinates by 
\bea
z_0&=& H^{-1}\sinh(H\tau)\nonumber\\
z_i&=&H^{-1} \cosh(H\tau)\,\sin\chi\,\omega_i \ \ \ \ \ 
\mbox{with}\; i=1,2,3\ \ \mbox{and}\ \ \sum_i\omega_i^2=1 
\nonumber\\
z_4&=& H^{-1}\cosh(H\tau)\,\cos\chi\ .
\label{hyperspherical}
\eea
These coordinates cover the whole manifold,
aside from a set of measure zero.
The metric has the form (\ref{deSpos}), with $r=\sin\chi$.
If we now define $\cosh(H\tau)=1/\cos\rho$, 
with 
$-\frac{\pi}{2}\leq\rho\leq\frac{\pi}{2}$,
so that 
\be
z_0 = H^{-1}\tan\rho\ ;\qquad
z_4 = H^{-1}  \cos \chi/\cos\rho\ ,
\ee
the metric takes the form
\be
ds^2=(H\cos\rho)^{-2}\left[-d\rho^2+d\chi^2+\sin^2\chi d\Omega_2^2
\right]\ ,
\ee
Fixing the spherical coordinates,
it corresponds to a finite square of side $\pi$,
which is the Penrose diagram for this space, see Fig.1.

We will next consider four different choices for the
one-form $X$, which are naturally associated to
four different coordinate systems:
the three FRW forms (\ref{deSflat},\ref{deSpos},\ref{deSneg})
and static coordinates.
Throughout this discussion it is important to keep in mind
that the analytic continuation of the metric
only depends on $X$ and not on the coordinate system:
It is just easier to describe if we choose a suitable 
coordinate system.
We will make this point clear by also giving the form
of $X$ in the global coordinates $\rho,\chi$ of the Penrose diagram.

\begin{SCfigure}
\includegraphics[scale=1]{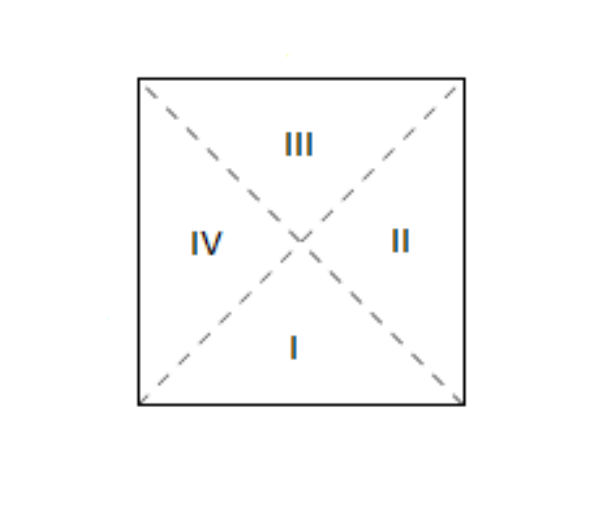}
\caption{\small Penrose diagram of de Sitter space.
The coordinates are $\rho$ (timelike, vertical)
and $\chi$ (spacelike, horizontal).
Every point in the interior of the square
corresponds to a 2-sphere,
while the left and right edges correspond to the poles $\chi=0,\pi$.}
\label{fig:1}
\end{SCfigure}

\subsection{First choice of X}

We start from the hyperspherical coordinates (\ref{hyperspherical}),
which cover the whole de Sitter space.
They are related to the FLRW coordinates of (\ref{deSpos})
by $r=\sin\chi$.
The surfaces of constant $\tau$ define an ADM foliation with $\Sigma=S^3$.

Let us choose the one-form $X=d\tau$.
The corresponding vectorfield $X^\mu$
has components $(1,0,\ldots,0)$ in this coordinate system.
The analytically continued metrics are
\begin{equation}
ds^2(\sigma)= 
\sigma d\tau^2
+\frac{1}{H^2}\cosh^2(H\tau) (d\chi^2+\sin^2\chi d\Omega^2_2)\ . 
\label{desv1}
\end{equation}
This choice has the virtue that $X$,
and therefore also $g(\sigma)$, are defined globally.
In particular, for $\sigma=1$ we obtain a global
Euclidean metric.
However, the Ricci tensor has the form
\begin{equation}
R_{(\sigma)\mu\nu} 
= -\frac{3H^2}{\sigma} g_{(\sigma)\mu\nu}
+2H^2\frac{(1+\sigma)}{\sigma}P_{\mu\nu}  
\label{riccinot}
\end{equation}
where $P_{\mu\nu}$ is the projector on the spacelike hypersurfaces.
This means that for $\sigma>-1$ these metrics are not Einstein.
A fortiori they cannot be maximally symmetric.
An examination of the Killing equation shows that
only the generators of the group $SO(d)$
of isometries of the constant time surfaces
are Killing vectors for all $\sigma$.
All the other vectorfields that are Killing for
$\sigma=-1$ are not Killing for $\sigma>-1$.
We can understand this by observing that,
unlike the Minkowski case,
a change of $\sigma$ cannot be absorbed in a rescaling of $\tau$.\\


\subsection{Second choice of $X$}

Next consider the FRW coordinates where $\Sigma=\mathbb{H}^d$
is a space of constant negative curvature.
The metric has the form (\ref{deSneg}),
but we replace the coordinate $r$ by $\chi$,
defined by $r=\sinh\chi$.
We choose $X=d\bar\tau$ in these coordinates.
Then 
$$
ds^2(\sigma)= \sigma d\bar{\tau}^2+\frac{1}{H^2}\sinh^2 H \bar{\tau}\; (d\chi^2+\sinh^2\chi d\Omega_2^2)\ .
$$
As in the positively curved case, the Ricci tensor is given
by (\ref{riccinot}), so 
these metrics are not Einstein for $\sigma>-1$.
Since $\bar{\tau} = H^{-1} \mbox{arcosh}\,\left(H z_4 \right) = H^{-1}\mbox{arcosh}\,\left(\cos \chi \cosh H\tau \right)$, 
the vectorfield $X^\#$ (with components $X^\mu$)
reads in global coordinates:
\be
X^\#=\partial_{\bar{\tau}} =  \frac{H \cos^2 \rho }{ \sqrt{ \cos^2 \chi - \cos^2 \rho } } \left( \tan \rho \cos \chi \, \partial_{\rho} +  \sin \chi  \; \partial_{\chi} \right)
\label{second}
\ee
This vector is defined only in a region of the de Sitter space which satisfies:
\be
\cos^2 \chi > \cos^2 \rho \iff z_4 ^2 > H^{-2}
\ee
Wherever it is well-defined, its norm is equal to $-1$

{\it
This vectorfield becomes singular on the hypersurface $z_4^2=H^{-2}$, which is equivalent to $\bar{\tau}=0$.
The singularity corresponds to the diagonals in
the Penrose diagram.
The vectorfield is imaginary in the quadrants III and IV.
}

\begin{SCfigure}
\includegraphics[scale=0.3]{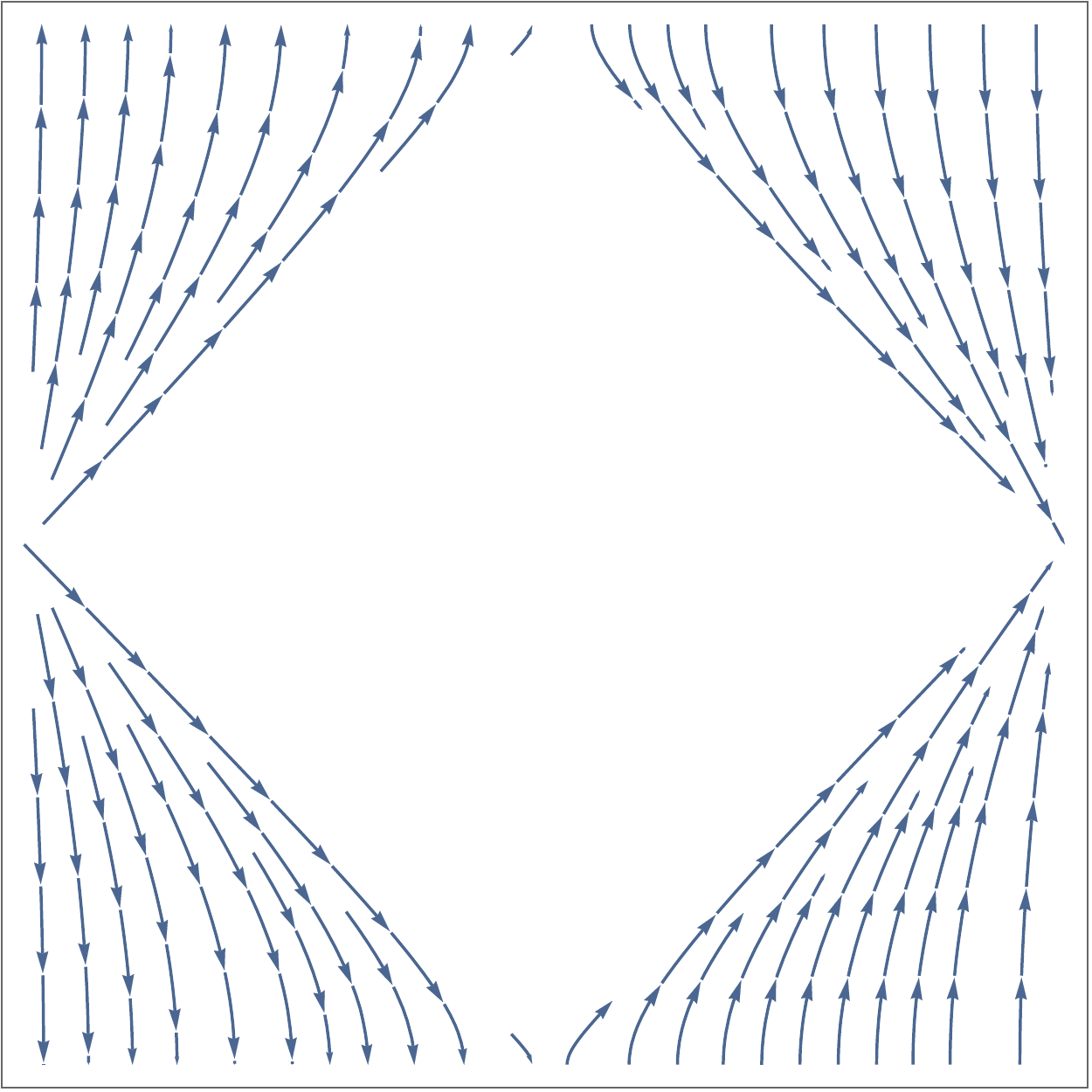}
\caption{\small The second choice for the vectorfield $X^\mu$.
It is imaginary in the central diamond.}
\label{fig:1}
\end{SCfigure}

\subsection{Third choice of $X$}

Now we come to the FRW coordinates with flat spatial sections,
where the metric has the form (\ref{deSflat}).
Once again we choose $X_\mu=(1,0,\ldots 0)$.
The analytically continued metric is
\be
ds^2(\sigma)= \sigma dt^2 + \frac{1}{H^2}e^{2Ht}\sum_{i=1}^{3}dx_i^2 \ .
\ee
and the corresponding Riemann tensor is (in any dimension):
\be
R_{(\sigma)\mu \nu \rho \sigma} = 
-\frac{H^2}{\sigma}  
\left[g_{(\sigma)\mu\rho}\,g_{(\sigma)\nu\sigma} - g_{(\sigma)\mu\sigma}\,g_{(\sigma)\nu\rho}\right]\ .
\ee
Thus the metric is maximally symmetric for all $\sigma$.
Indeed, the following vectors are Killing:
\begin{equation}
\begin{split}
    & P_i=\partial_i\ , \hspace{1 cm} 
    M_{ij}=x_i\partial_j-x_j \partial_j
    \ , \hspace{1 cm} B=-\frac{1}{H}\partial_t + \sum_{k=1}^{3}x_k\partial_k\\
    & K_i=-\frac{1}{H}x_i\partial_t
    +\frac{1}{2} \left[-\sigma\, e^{-2Ht}
    -\sum_{k=1}^{3}x_k^2\right] \partial_i + x_i \sum_{k = 1}^{3} x_k \partial_k
\end{split}
\label{killplalam}
\end{equation}
and satisfy {\it the same algebra}  for all $\sigma$:
\begin{equation*}
\begin{split}
    & \left[ P_{i}  , B \right] = P_i \hspace{5 cm} \left[ P_{i}  , K_j \right] = \delta_{i j} B - M_{i j} \\
    & \left[ M_{i j}  , B \right] = 0 \hspace{5 cm} \left[ M_{i j} , K_p \right] = \delta_{j p} K_i - \delta_{i p} K_j\\
    & \left[ B  , K_i \right] = K_i  \hspace{5 cm} \left[ K_i  , K_j \right] = 0
\end{split}
\end{equation*}

\begin{figure}
\begin{center}
\includegraphics[scale=0.3]{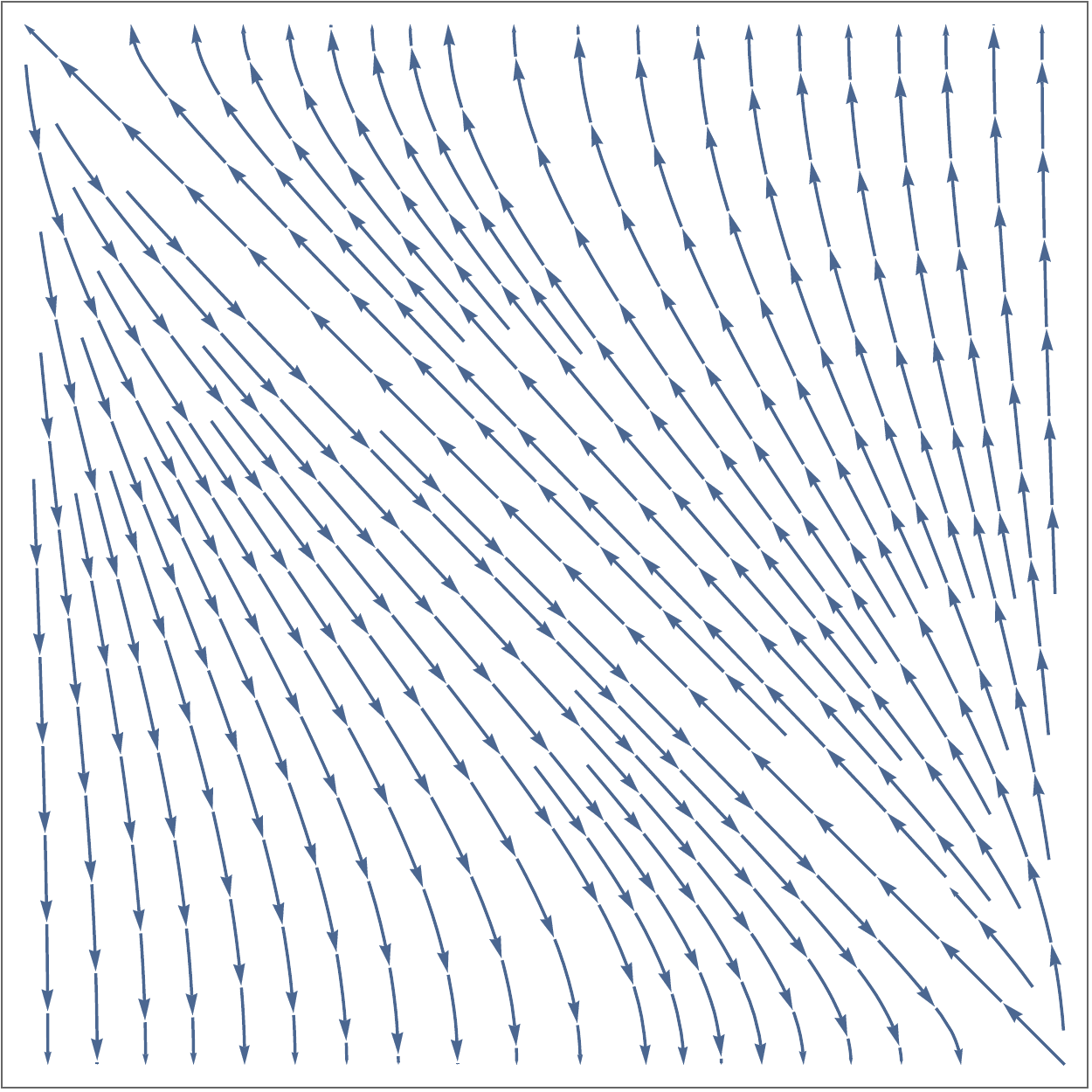}\qquad
\includegraphics[scale=0.3]{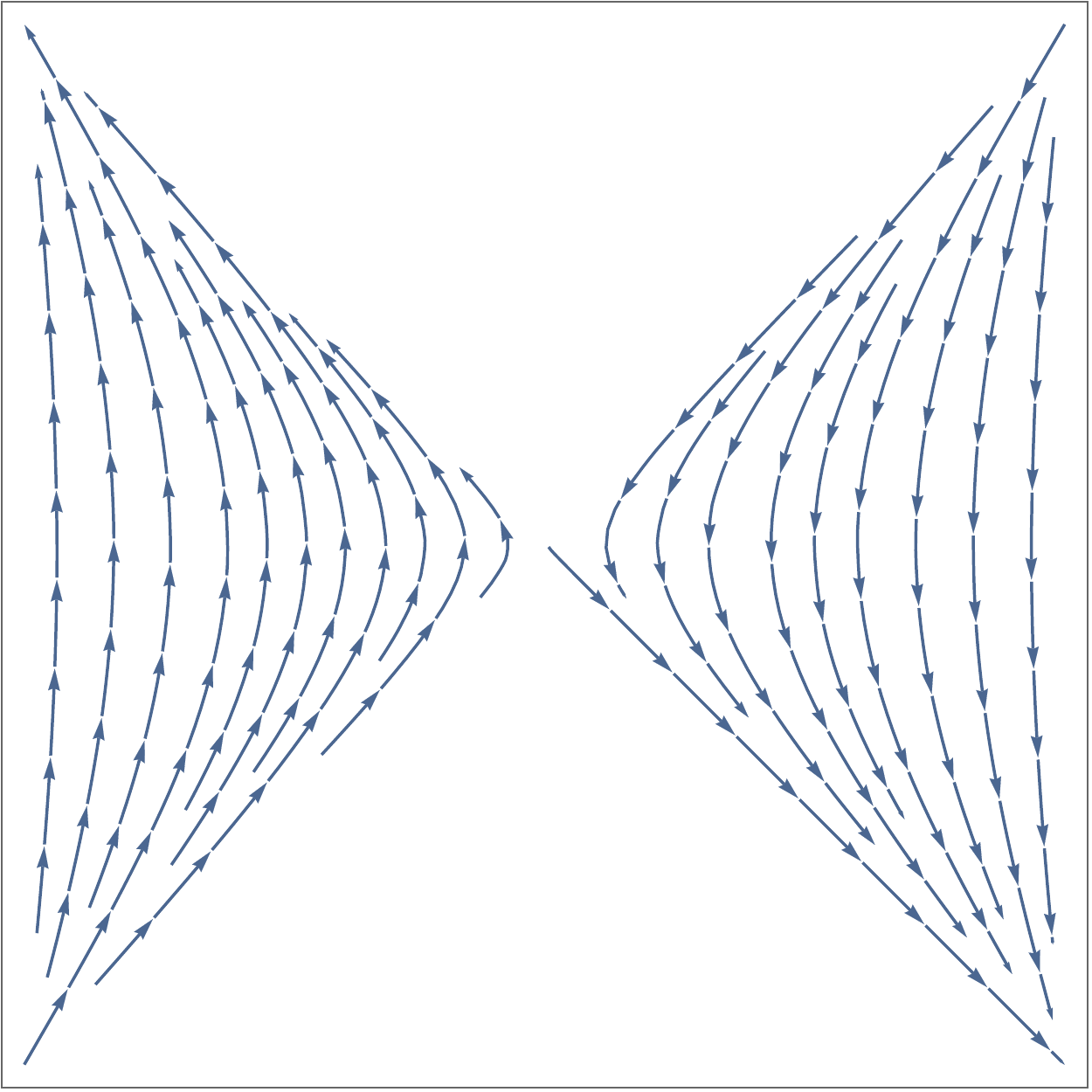}
\caption{\small Left: The third choice for the vectorfield $X^\mu$.
It is singular on the diagonal (the horizon of the observer
at the north pole).
Right: The fourth choice for the vectorfield $X^\mu$.
It is imaginary in regions I and III.}
\label{fig:1}
\end{center}
\end{figure}

Since $t = (1/H)\ln\left[ z_0 - z_4 \right] =
(1/H) \ln\left[ \sinh \tau - \cos \chi \cosh \tau \right] $, 
the vectorfield $X^\#$ can be expressed 
as follows in global coordinates:
\be
X^\# = \partial_t = H \cos \rho \; \frac{1 -\cos \chi \sin \rho }{ \sin \rho - \cos \chi } \left[ \partial_\rho +  \frac{\sin \chi \cos \rho }{ \cos \chi \sin \rho - 1 } \; \partial_{\chi} \right]
\label{third}
\ee
It becomes singular for $\sin \rho = \cos \chi $, 
which is equivalent to $z_0=z_4$: 
this is true for $t\to-\infty$ 
and it means that the vector field $X$ 
is not well-defined on the boundary
of the region I$\cup$IV, in the Penrose diagram in Fig.\ref{fig:1}.
Thus the domain of definition of the analytically continued metric
is one half of de Sitter space.

\subsection{Fourth choice of $X$}

The static coordinates on de Sitter space are defined by:
\bea
z_0&=&H^{-1} \cos\zeta\sinh t\nonumber\\
z_i&=&H^{-1} \sin\zeta\;\omega_i \ \ \ \ \ 
\mbox{with}\; i=1,2,3\ \ \mbox{and}\ \ \sum_i\omega_i^2=1 
\nonumber\\
z_4&=&H^{-1} \cos\zeta\cosh t
\eea
where $t \in (-\infty;+\infty)$ and $\zeta\in(-\pi/2;\pi/2)$.
Choosing $X=\cos\sigma\, dt$
we get the following family of metrics:
\begin{equation}
ds^2(\sigma)=H^{-2} \left[\sigma \cos^2 \zeta \;dt^2+  d\zeta^2+\sin^2\zeta\;d\Omega^2_{2} \right]\ . 
\end{equation}

The coordinate $\zeta$ is related to the coordinates 
of the Penrose diagram by:
$\sin\zeta = \sin \chi/\cos\rho$.
Then, the vector $X^\#$ can be expressed in global coordinates:
\be
X^\# = H\partial_t =\frac{H \cos^2 \rho}{ \sqrt{\cos^2 \chi -\sin^2 \rho} } \left[ \cos \chi \; \partial_{\rho} - \sin \chi \tan \rho \; \partial_{\chi} \right]
\label{fourth}
\ee
Its norm is equal to $-1$, but this vector is defined only in a region of the de Sitter space which satisfies:
\be
\cos^2 \chi > \sin^2 \rho \iff z_0 ^2 < z_4 ^2
\ee
This is in contrast to Anti-de Sitter space,
where the static coordinates cover the whole manifold.

The Riemann tensor is:
\be
R_{(\sigma)\mu \nu \rho \sigma} = 
H^2  
\left[g_{(\sigma)\mu\rho}\,g_{(\sigma)\nu\sigma} - g_{(\sigma)\mu\sigma}\,g_{(\sigma)\nu\rho}\right]
\ee
so, in this case too, 
the metric is maximally symmetric for all $\sigma$.
One can deform the Killing vectors of the de Sitter group
with the parameter $\sigma$
in such a way that their algebra remains unchanged
for $-1\leq\sigma<0$.
However, for $\sigma>0$ they satisfy the algebra of $SO(5)$.

\subsection{General result}

From the preceding examples one may suspect that there exists no
globally defined normalized timelike one-form $X_\mu$ 
such that the analytically continued metrics are maximally symmetric.
Let us formulate the problem precisely.
Suppose that the Lorentzian metric $g_{\mu\nu}$ is maximally symmetric.
For convenience, let $\lambda=\sigma+1$ be infinitesimal.
The original Lorentzian metric corresponds to $\lambda=0$.
For an infinitesimal $\lambda$, 
$\delta g_{\mu\nu}=g_{(\lambda)\mu\nu}-g_{\mu\nu}
=\lambda X_\mu X_\nu$.
If $g_{(\lambda)}$ and $g$ are both maximally symmetric,
then there exists an infinitesimal conformal isometry, 
\be
\delta g_{\mu\nu}=\lambda\left(\nabla_\mu W_\nu+\nabla_\nu W_\mu
- c g_{\mu\nu}\right)\ ,
\label{agata}
\ee
for some vectorfield $W$ and constant $c$.
Conversely, it is shown in Appendix A that if 
$g$ is maximally symmetric and (\ref{agata}) holds, 
then $g_{(\lambda)}$ is also maximally symmetric.
The constant $c$ is related to the constant $f(\lambda)$
of (\ref{alessio}) by $f(\lambda)=1-c\lambda+O(\lambda^2)$.

Therefore, a local necessary and sufficient condition
for  the metric $g_{(\lambda)}$ to be maximally symmetric,
is that there exists a vectorfield $W$
and a constant $c$ such that
\be
\nabla_\mu W_\nu+\nabla_\nu W_\mu
-c g_{\mu\nu}=X_\mu X_\nu\ .
\label{ennio}
\ee

Let us write (\ref{ennio}) explicitly
in the coordinate system of the Penrose diagram.
\bea
2\partial_\rho W_\rho-2\tan\rho W_\rho+c\sec^2\rho&=&X_\rho^2
\label{con00}\\
\partial_\rho W_\chi+\partial_\chi W_\rho
-2\tan\rho W_\chi&=&X_\rho X_\chi
\label{con01}\\
\partial_\rho W_\theta+\partial_\theta W_\rho
-2\tan\rho W_\theta&=&X_\rho X_\theta
\label{con02}\\
\partial_\rho W_\phi+\partial_\phi W_\rho
-2\tan\rho W_\phi&=&X_\rho X_\phi
\label{con03}\\
2\partial_\chi W_\chi-2\tan\rho W_\rho
-c\sec^2\rho&=&X_\chi^2
\label{con11}\\
\partial_\chi W_\theta+\partial_\theta W_\chi
-2\cot\chi W_\theta&=&X_\chi X_\theta
\label{con12}\\
\partial_\chi W_\phi+\partial_\phi W_\chi
-2\cot\chi W_\phi&=&X_\chi X_\phi
\label{con13}\\
2\partial_\theta W_\theta-(2\tan\rho W_\rho
-2\cot\chi W_\chi
+c\sec^2\rho)\sin^2\chi
&=&X_\theta^2
\label{con22}\\
\partial_\theta W_\phi+\partial_\phi W_\theta
-2\cot\theta W_\phi&=&X_\theta X_\phi
\label{con23}\\
2\partial_\phi W_\phi
+2\sin\theta\cos\theta W_\theta
-(2\tan\rho W_\rho
-2\cot\chi W_\chi
+c\sec^2\rho)\sin^2\chi\sin^2\theta
&=&X_\phi^2\qquad\qquad
\label{con33}
\eea

We already have two solutions of these equations:
they are given by the vectorfields (\ref{third},\ref{fourth}),
together with the corresponding infinitesimal isometries and rescalings:
\bea
W_{flat} &=& -\frac{1}{2} H \cos \rho \; \frac{1 -\cos \chi \sin \rho }{ \sin \rho - \cos \chi } \left[ \partial_\rho +  \frac{\sin \chi \cos \rho }{ \cos \chi \sin \rho - 1 } \; \partial_{\chi} \right] 
\nonumber\\
X_{flat} &=& H \cos \rho \; \frac{1 -\cos \chi \sin \rho }{ \sin \rho - \cos \chi } \left[ \partial_\rho +  \frac{\sin \chi \cos \rho }{ \cos \chi \sin \rho - 1 } \; \partial_{\chi} \right]
\label{xflat}
\\
c&=&-1
\nonumber
\eea
and
\bea
W_{sta} &=& -\frac{1}{2} \arctanh\left(\frac{\sin\rho}{\cos\chi}\right)
\frac{H \cos^2 \rho}{ \sqrt{\cos^2 \chi -\sin^2 \rho} } \left[ \cos \chi \; \partial_{\rho} - \sin \chi \tan \rho \; \partial_{\chi} \right]
\nonumber\\
X_{sta} &=& \frac{H \cos^2 \rho}{ \sqrt{\cos^2 \chi -\sin^2 \rho} } \left[ \cos \chi \; \partial_{\rho} - \sin \chi \tan \rho \; \partial_{\chi} \right]
\label{xstat}
\\
c&=&0
\nonumber
\eea

As we have discussed earlier, these solutions are singular
and we would like to prove in general that the equations cannot have
regular solutions.
We have not been able to do so in full generality.
However, we can make definite statements
when we linearize the equations around one of the
two solutions given above.
Denote $\overline W_\mu$, $\overline X_\mu$ a solution of the full equations
and write
$$
W_\mu=\overline W_\mu+\delta W_\mu\ ;\qquad
X_\mu=\overline X_\mu+\delta X_\mu\ .
$$

We show in Appendix B that when $\overline X_\mu$ is the vectorfield
(\ref{xflat}) (the one related to the flat FLRW slicing), 
the linearized equations have no real solutions.
Thus the third vectorfield is an isolated solutions of 
the system (\ref{con00}-\ref{con33}).
On the other hand, when $\overline X_\mu$ is the vectorfield
(\ref{xstat}),
(the one related to static coordinates), there is a family of
solutions of the linearized equations, 
but the perturbed solutions are all singular.
Thus, at the linearized level, we could indeed prove that there
are no globally regular solutions of (\ref{con00}-\ref{con33}).

\section{Schwarzschild spacetime}

As an example of a non-maximally symmetric spacetime we consider here
Schwarzschild spacetime. It has four Killing vectors
generating the isometry group
$SO(3)\times T$, where $T$ denotes time translations.

\subsection{First choice of $X$}

We use Schwarzschild coordinates.
Choosing $X=\sqrt{1-\frac{2M}{r}}dt$, 
the analytically continued metric is
\be
ds^2 = \sigma \left(1 - \frac{2M}{r} \right)dt^2 
+ \left(1 - \frac{2M}{r} \right)^{-1}dr^2 + r^2 d\Omega_{2}^2\ .
\label{eucs}
\ee
This metric is Ricci-flat for all $\sigma$
and all the Killing vectors of the Schwarzschild metric
are Killing vectors for all $\sigma$.
However, this analytic continuation is not globally defined.

\subsection{Second choice of $X$}

Alternatively, let us try to perform the analytic continuation
at the level of Kruskal coordinates:
\be
ds^2(\sigma) = \frac{16M^2}{X^2 -T^2 } \frac{W(z)}{W(z)+1}
\left[ \sigma dT^2 + dX^2 \right] + 4M^2 \left(W(z)+1 \right)^2 d\Omega_{2}^2 
    \label{genekru}
\ee
where $z \equiv\frac{X^2-T^2}{e}$ and $W$ is the Lambert function. The Ricci tensor has a complicated, non-vanishing expression, with a prefactor $1+\sigma$.
Thus, the analytically continued metric does not satisfy
Einstein's equations in vacuum, even for an infinitesimal
deformation.

Concerning the symmetries, we find that the generators of $SO(3)$
are preserved, but the timelike vector
$$
K_t=4 M \partial_t=X \partial_T+T\partial_X
$$ 
is a Killing vector only for $\sigma=-1$.

This should be compared to the standard analytic continuation
of the Schwarzschild metric, based on the replacement
$T\to -iT$ in the Lorentzian Kruskal metric \cite{Hawking:1976jb}.
The difference is that whereas with the present definition
one only changes the sign of the $dT^2$ term, keeping all the rest
unchanged, in the Cambridge definition one also changes 
$z$ to $X^2+T^2$,
The resulting Euclidean metric is still a solution of the
vacuum Einstein equations and still has all the Killing vectors.
However, it is only defined for $r>2M$.
In fact, one can transform it back to Schwarzschild coordinates
and then it coincides with (\ref{eucs}) for $\sigma=1$.
Thus, the Cambridge definition of Euclidean Schwarzschild
metric is equivalent to
the continuation based on our first choice of $X$.

\section{Discussion}

The Wick rotation is a problematic notion when gravity is involved,
or more generally when spacetime is curved.
When interpreted as a continuation of some time coordinate,
and for a fixed background metric,
it has ambiguities that are hard to settle.
Things are worse when gravity is dynamical.
The Euclidean Quantum gravity programme simply assumed that the
functional integral should be performed on all Euclidean
metrics. 
In practice, this has led to many useful and deep insights,
but it faces the issue of the classification of
all topologies, which is unsolvable in four dimensions.

The alternative notion of continuing the metric seems to be better.
In particular, it has the attractive feature that
the Wick-rotated metric are defined on the same manifold.
If a sum over topologies is needed, it is restricted to
manifolds admitting a nowhere vanishing vectorfield,
which is a much tamer set.
It has been seen from numerical simulations with CDTs
that the restriction to triangulations admitting a Lorentz metric
has a very beneficial effect on the path integral.

We have seen here that in certain important cases,
the requirement of keeping the spacetime manifold fixed
during the Wick rotation
clashes with other desirable properties,
such as sending local solutions of Einstein's equations to
other local solutions of the same equations (possibly up to a 
change of sign of the cosmological constant)
and/or preserving the number of Killing vectors.
With the definition of Wick rotation given in Section 1.3,
and for de Sitter and Schwarzschild spacetimes,
we have seen that when the vectorfield $X^\mu$ is such that the
Euclidean metric solves the field equations locally,
the solution does not extend to the whole manifold.
This is somewhat analogous to the behavior of other fields
under Wick rotation, as we have already observed in section 1.5.
At least for the cases that we have discussed, this behavior
is clearly related to the presence of horizons.
It is a familiar fact in Euclidean quantum gravity, 
when the Wick rotation is
interpreted as a complexification of the coordinates,
that the region beyond the horizon disappears
in the Euclidean section \cite{Gibbons:1976ue}.
We see that the same is true also with the alternative
definitions of Wick rotation discussed here.

\bigskip

{\bf Acknowledgments.} We would like to thank M. Visser
and C. Wetterich for useful discussions.

\goodbreak

\begin{appendix}

\section{Wick rotation, Einstein equations and Killing vectors}

One can ask what are the conditions for an analytically
continued metric of the form (\ref{ancont}) to 
maintain a constant number of Killing vectors,
as the parameter $\sigma$ varies continuously.
We address this question for infinitesimal deformations
of the metric. It is then more convenient to use the parameter 
$\lambda=\sigma+1$, so that the initial Lorentzian metric
corresponds to $\lambda=0$.
Obviously, a sufficient condition is that the deformed metric 
is related to the original metric by an isometry.
We prove here a slightly more general result,
which covers the examples of section 3.
\medskip

\noindent
{\bf Proposition 1.}
Suppose that there exists a vectorfield $W$ and a constant 
$f(\lambda)$, depending on $\lambda$, such that 
\be
g_\lambda=f(\lambda)g+\lambda\Lie_W g\ .
\label{alessio}
\ee
If the metric $g$ is Einstein, with
$$
Ric(g)=\Lambda g
$$
($Ric(g)$ denoting the Ricci tensor of $g$)
then $g_\lambda$ is Einstein with 
$$
Ric(g_\lambda)=\frac{\Lambda}{f(\lambda)} g_\lambda\ .
$$
\smallskip

\noindent
{\it Proof.} This follows immediately from the fact that $g$ and $g_\lambda$
are isometric up to a constant rescaling.
\medskip

\noindent
{\bf Proposition 2.}
Suppose that there exists a vectorfield $W$ and a constant 
$f(\lambda)=1+O(\lambda)$, such that (\ref{alessio}) holds.
Then if $K_i$ are Killing vectors for $g$ satisfying the algebra
\be
[K_i,K_j]=f_{ij}{}^k K_k\ ,
\label{algebra}
\ee
to first order in $\lambda$ the vectorfields
\begin{equation}
K_i(\lambda) = K_i+\lambda[W,K_i] 
\label{killamb}
\end{equation}
are Killing vectors for $g_\lambda$ and obey the same algebra
(\ref{algebra}).
\smallskip

\noindent
{\it Proof.} For each of the vectorfields $K_i$,
suppressing the index $i$,
$$
\Lie_K g_\lambda=\lambda/f\Lie_K\Lie_W g
=\lambda/f[\Lie_K,\Lie_W]g
=\lambda[\Lie_K,\Lie_W]g_\lambda+O(\lambda^2)\ .
$$
To first order in $\lambda$ we therefore have
$$
\Lie_{K+\lambda[W,K]}g_\lambda=0\ ,
$$
showing that $K(\lambda)$ is a Killing vector of $g_\lambda$.

To first order in $\lambda$, using the Jacobi identity one gets
$$
[K_i(\lambda),K_j(\lambda)]=f_{ij}{}^k K_k
+\lambda\left([K_i,[W,K_j]]+[K_j,[K_i,W]]\right)
=f_{ij}{}^k K_k(\lambda)\ ,
$$
so the algebra is unchanged, QED.

\section{Solutions of linearized equations}

Here we prove the statements made in the end of section 3.5
on the solutions of the linearized equations.
Exploiting the fact that $\overline X_\theta=\overline X_\phi=0$,
the linearized equations read
\bea
2\partial_\rho \delta W_\rho-2\tan\rho \delta W_\rho+\delta c\sec^2\rho&=&2 \overline X_\rho \delta X_\rho
\label{clin00}\\
\partial_\rho \delta W_\chi+\partial_\chi \delta W_\rho
-2\tan\rho \delta W_\chi&=&\overline X_\rho \delta X_\chi + \overline X_\chi \delta X_\rho
\label{clin01}\\
\partial_\rho \delta W_\theta+\partial_\theta \delta W_\rho
-2\tan\rho \delta W_\theta&=&\overline X_\rho \delta X_\theta
\label{clin02}\\
\partial_\rho \delta W_\phi+\partial_\phi \delta W_\rho
-2\tan\rho \delta W_\phi&=&\overline X_\rho \delta X_\phi
\label{clin03}\\
2\partial_\chi \delta W_\chi-2\tan\rho \delta W_\rho
-\delta c\sec^2\rho&=&2 \overline X_\chi \delta X_\chi
\label{clin11}\\
\partial_\chi \delta W_\theta+\partial_\theta \delta W_\chi
-2\cot\chi \delta W_\theta&=&\overline X_\chi \delta X_\theta
\label{clin12}\\
\partial_\chi \delta W_\phi+\partial_\phi \delta W_\chi
-2\cot\chi \delta W_\phi&=&\overline X_\chi \delta X_\phi
\label{clin13}\\
2\partial_\theta \delta W_\theta-(2\tan\rho \delta W_\rho
-2\cot\chi \delta W_\chi
+\delta c\sec^2\rho)\sin^2\chi
&=&0
\label{clin22}\\
\partial_\theta \delta W_\phi+\partial_\phi \delta W_\theta
-2\cot\theta \delta W_\phi&=&0
\label{clin23}\\
2\partial_\phi \delta W_\phi
+2\sin\theta\cos\theta \delta W_\theta
\qquad\qquad\qquad\qquad
\qquad\qquad
&&
\nonumber\\
-(2\tan\rho \delta W_\rho
-2\cot\chi \delta W_\chi
+\delta c\sec^2\rho)\sin^2\chi\sin^2\theta
&=&0\qquad\qquad
\label{clin33}
\eea
This is a linear system of equations for $\delta W_\mu$
and $\delta X_\mu$.
It is to be supplemented by the condition,
\be
\bar g^{\mu\nu}\overline X_\mu\delta X_\nu=0\ ,
\ee
which follows from the normalization of $X_\mu$.

One can solve algebraically the four equations
(\ref{clin00},\ref{clin11},\ref{clin02},\ref{clin03})
to express all the $\delta X_\mu$
as linear functions of $\overline X_\mu$, the coordinates,
$\delta W_\mu$ and their derivatives.
These solutions can be substituted in the remaining six equations
obtaining a linear system for the $\delta W_\mu$ alone.
Using the normalization $\overline X_\chi^2-\overline X_\rho^2=1$,
the resulting equations read:
\bea
\left[ \partial_\chi + \frac{\overline X_\rho}{ \overline X_\chi} \tan \rho - \frac{\overline X_\chi}{\overline X_\rho} \left( \partial_\rho- \tan \rho \right) \right] \delta W_\rho  +\left[ \partial_\rho -2\tan \rho - \frac{\overline X_\rho}{ \overline X_\chi} \partial_\chi\right]\delta W_\chi &=& -\frac{\delta c }{2 \overline X_\rho \overline X_\chi}  
\label{lin1}\\
\partial_\theta \delta W_\rho-\frac{\overline X_\rho}{\overline X_\chi} \partial_\theta \delta W_\chi+\left[ \partial_\rho 
-2\tan\rho  -\frac{\overline X_\rho}{\overline X_\chi} \left( \partial_\chi 
-2\cot\chi \right) \right] \delta W_\theta  &=&0
\label{lin2}\\
\partial_\phi \delta W_\rho-\frac{\overline X_\rho}{\overline X_\chi} \partial_\phi \delta W_\chi+\left[ \partial_\rho 
-2\tan\rho  -\frac{\overline X_\rho}{\overline X_\chi} \left( \partial_\chi 
-2\cot\chi \right) \right] \delta W_\phi  &=&0
\label{lin3}\\
 2\tan\rho \delta W_\rho
-2\cot\chi \delta W_\chi 
-2\sin^{-2}\chi\partial_\theta \delta W_\theta
&=& -\delta c \sec^2\rho
\label{lin4}\\
\partial_\phi \delta W_\theta + 
\left(\partial_\theta -2\cot\theta \right) \delta W_\phi&=&0
\label{lin5}\\
\sin\theta\left( \cos\theta -\sin \theta \partial_\theta \right) \delta W_\theta+ \partial_\phi \delta W_\phi
&=&0\qquad\qquad
\label{lin6}
\eea

The natural way of solving these equations
is by separation of variables:
$$
\delta W_\mu= A_\mu(\rho,\chi)G_\mu (\theta,\phi)\ .
$$
We proceed under this assumption,
considering first the linearization around (\ref{xstat})
and then the linearization around (\ref{xflat}).

\subsection{Static case}

\textit{First possibility:  $\delta c \ne 0 $}\\
\smallskip

From equation (\ref{lin1}),
since the r.h.s. is a function of $\rho$ and $\chi$ only,
$G_\rho$ and $G_\chi$ must be constants which,
up to rescalings of $A_\rho$ and $A_\chi$ we can assume equal to one.
From (\ref{lin4}), $G_\theta$ must be a linear function of $\theta$,
which would be strange, considering that everything
is expressed as trigonometric functions. 
Indeed, using this property in equations (\ref{lin5},\ref{lin6})
we conclude that $G_\phi$ should be a linear function of $\phi$.
This is inconsistent with periodicity, leading to $G_\phi=0$.
Then, the same equations imply that also $G_\theta=0$.

So equations (\ref{lin1},\ref{lin4}) simplify:
\bea
\left[ \partial_\chi + \frac{\overline X_\rho}{ \overline X_\chi} \tan \rho - \frac{\overline X_\chi}{\overline X_\rho} \left( \partial_\rho- \tan \rho \right) \right] A_\rho  +\left[ \partial_\rho -2\tan \rho - \frac{\overline X_\rho}{ \overline X_\chi} \partial_\chi\right] A_\chi &=& -\frac{\delta c }{2 \overline X_\rho \overline X_\chi} 
\label{deltac1}\\
\tan\rho A_\rho-\cot\chi A_\chi&=&  -\frac{\delta c}{2} \sec^2\rho
\label{deltac2}
\eea
Inserting the algebraic constraint (\ref{deltac2}) into equation (\ref{deltac1}) we obtain $\delta c =0$ and so we get no solution. \\

\bigskip

\textit{Second possibility: $\delta c = 0 $ and $\partial_\theta \delta W_\theta = 0$ }
\smallskip

Equations (\ref{lin5},\ref{lin6}) can be solved in one of three ways:
either $A_\phi=0$ and $A_\theta=$const,
or $A_\phi=$const and $A_\theta=0$ 
or $A_\phi=A_\theta$.

In the first case from (\ref{lin6}) we have $G_\theta=0$
and therefore $\delta W_\theta=\delta W_\phi=0$.
In the second case  equations (\ref{lin5},\ref{lin6})
have the solution $\delta W_\theta=0$
and $G_\phi=a \sin^2\theta$, where $a$ is a constant.

In the third case (\ref{lin5},\ref{lin6}) become
\bea
\partial_\phi G_\theta + 
\left(\partial_\theta -2\cot\theta \right) G_\phi&=&0\\
\sin\theta \cos\theta G_\theta+ \partial_\phi G_\phi
&=&0
\eea
Since we are looking for $G$'s periodic in $\phi$ we can write:
$$
G_\theta = \sum_m c^m_\theta e^{im\phi} \;\qquad
G_\phi = \sum_m c^m_\phi (\theta) e^{im\phi}
$$
And so:
\bea
\partial_\theta c^m_\phi -2\cot\theta c^m_\phi&=&-im c^m_\theta \\
-\sin \theta \cos\theta c^m_\theta &=&im c^m_\phi
\eea
If $m\ne 0$ and $m\ne 1$, $c^m_\theta = c^m_\phi =0$.\\
If $m=1$, $c^1_\theta= a_1$ and $c^1_\phi= i a_1 \sin \theta \cos \theta$ and for reality condition $a_1 =0$.\\
If $m=0$, $c^0_\theta=0$ and $c^0_\phi= a_2 \sin^2 \theta$.\\
So in all three cases equations (\ref{lin5},\ref{lin6})
imply that
$\delta W_\theta =0$ and $\delta W_\phi=A_\phi \sin^2 \theta$.

Now let us come to equation (\ref{lin1}).
It can be solved in one of three ways:
either $G_\rho=0$ and $G_\chi=$const,
or $G_\rho=$const and $G_\chi=0$,
or $G_\rho=G_\chi$.
In the first two cases equation (\ref{lin4}) gives
$\delta W_\rho=\delta W_\chi=0$.
Therefore in the following we consider only the third case.

Using $G_\rho=G_\chi$ and the $\delta W_\theta=0$,
as derived above, equation (\ref{lin2})
is seen to be the derivative with respect to $\theta$ of (\ref{lin4}).

On the other hand doing the same in equation (\ref{lin3}),
the first two terms are seen to be 
the derivative with respect to $\phi$ of (\ref{lin4})
and therefore can be dropped.
The last term gives
$$
\left[ \partial_\rho 
-2\tan\rho  
-\frac{\overline X_\rho}{ \overline X_\chi} 
\left( \partial_\chi 
-2\cot\chi \right) \right]A_\phi=0
$$
where the prefactor of the second term in the square bracket
is $\overline X_\rho/\overline X_\chi=\cot\rho\cot\chi$.
The general solution of this equation is
$$
A_\phi=\sin^2\chi\sec^2\rho\, H\left(\frac{\cos\chi}{\sin\rho}\right)\ ,
$$
for some function $H$.

Putting $G_\rho=G_\chi$ in equations (\ref{lin1},\ref{lin4}), we obtain the following equations:
\bea
\left[ \partial_\chi + \frac{\overline X_\rho}{ \overline X_\chi} \tan \rho - \frac{\overline X_\chi}{\overline X_\rho} \left( \partial_\rho- \tan \rho \right) \right] A_\rho  +\left[ \partial_\rho -2\tan \rho - \frac{\overline X_\rho}{ \overline X_\chi} \partial_\chi\right] A_\chi &=& 0 
\label{secpos1}\\
\tan\rho A_\rho - \cot\chi A_\chi &=& 0 
\label{secpos2}
\eea
Solving the second equation for $A_\chi$ and substituting
in the first, one finds that it is identically satisfied.
Thus, the solution to the whole system is:
\bea
\delta W_\rho &=& A(\rho,\chi)G(\theta,\phi)\\
\delta W_\chi &=& \tan \rho \tan \chi A(\rho,\chi)G(\theta,\phi)\\
\delta W_\theta &=& 0 \\
\delta W_\phi &=& \sin^2\theta
\sin^2\chi\sec^2\rho\, H\left(\frac{\cos\chi}{\sin\rho}\right)
\eea
where $A(\rho,\chi)$ and $G(\theta,\phi)$ are arbitrary functions.

Now from equations (\ref{clin00},\ref{clin11}) 
we can determine:
\bea
\delta X_\rho &=& G(\theta,\phi)\frac{ \sqrt{\cos^2 \chi -\sin^2 \rho} }{H \cos \chi}\left[\partial_\rho A(\rho,\chi)-\tan\rho A(\rho,\chi) \right]\\
\delta X_\chi &=& G(\theta,\phi)\frac{ \sqrt{\cos^2 \chi -\sin^2 \rho} }{H \sin \chi} \left[ \partial_\chi \left( \tan \chi A(\rho,\chi)\right)- A(\rho,\chi) \right]
\eea
Imposing the normalization condition
\be
\overline X_\rho \delta X_\rho-\overline X_\chi \delta X_\chi=0
\label{norcon}
\ee
we obtain
$$
A(\rho,\chi) = \sec \rho \cos \chi\, F[\sec \rho \sin \chi]
$$
where $F(x)$ is a generic function, and therefore
\bea
\delta X_\rho &=& G(\theta,\phi)\frac{ \sqrt{\cos^2 \chi -\sin^2 \rho} }{H} \sec^2\rho \tan \rho \sin \chi \: F'[\sec \rho \sin \chi]\\
\delta X_\chi &=& G(\theta,\phi)\frac{ \sqrt{\cos^2 \chi -\sin^2 \rho} }{H } \sec^2\rho \cos \chi \:F'[\sec \rho \sin \chi]\\
\delta X_\theta &=& \frac{A}{\overline X_\rho}\partial_\theta G(\theta,\phi) \\
\delta X_\phi &=& \frac{A}{\overline X_\rho}\partial_\phi G(\theta,\phi)  \ .
\eea
So we have an infinite family of solutions.

We observe that these solutions are in general regular on
the horizon, so they cannot remove the singularity
of the solutions $\overline X_\mu$.

\bigskip

\textit{Third possibility: $\delta c = 0 $ and $\partial_\theta \delta W_\theta \ne 0$ }
\smallskip

From equation (\ref{lin1}), by the same argument
used for the second possibility, $G_\rho=G_\chi$.
Then (\ref{lin4}) implies that
$$
\sin^2\chi\tan\rho \frac{A_\rho}{  A_\theta}
-\sin\chi\cos\chi \frac{A_\chi}{  A_\theta}
= \frac{\partial_\theta G_\theta}{G_\rho}
$$
where we have separated the functional dependences on
$\rho$ and $\chi$ on the left and $\theta$ and $\phi$
on the right.
Therefore $G_\rho =\partial_\theta G_\theta (\theta,\phi)$
and we are left with the following equation:
$$
\tan\rho A_\rho-\cot\chi A_\chi -\sin^{-2}\chi A_\theta = 0
$$
then using equations (\ref{lin2},\ref{lin3}) we get:
\bea
\left[ A_\rho -\frac{\overline X_\rho}{\overline X_\chi} A_\chi\right] \partial^2_\theta G_\theta +G_\theta\left[ \partial_\rho 
-2\tan\rho  -\frac{\overline X_\rho}{\overline X_\chi} \left( \partial_\chi 
-2\cot\chi \right) \right] A_\theta  &=&0\\
\left[ A_\rho-\frac{\overline X_\rho}{\overline X_\chi} A_\chi \right] \partial_\phi \partial_\theta G_\theta+G_\phi\left[ \partial_\rho 
-2\tan\rho  -\frac{\overline X_\rho}{\overline X_\chi} \left( \partial_\chi 
-2\cot\chi \right) \right] A_\phi  &=&0
\eea
Separating variables as before this leads to
\bea
\partial^2_\theta G_\theta &=& a G_\theta \\
\partial_\phi \partial_\theta G_\theta &=& b G_\phi
\eea

Equations (\ref{lin5},\ref{lin6}) can be solved in one of three ways:
either $A_\phi=0$ and $A_\theta=$const,
or $A_\phi=$const and $A_\theta=0$ 
or $A_\phi=A_\theta$.

In the first case from (\ref{lin6}) we have $G_\theta=0$
and therefore $\delta W_\theta=\delta W_\phi=0$.
In the second case  equations (\ref{lin5},\ref{lin6})
have the solution $\delta W_\theta=0$
and $G_\phi=\sin^2\theta$.
This further implies $b=0$

In the third case (\ref{lin5},\ref{lin6}) become:
\bea
\partial_\phi G_\theta + 
\left(\partial_\theta -2\cot\theta \right) G_\phi&=&0\\
\sin\theta\left( \cos\theta -\sin \theta \partial_\theta \right) G_\theta+ \partial_\phi G_\phi &=&0
\eea
From the last two equations we get:
\be
\sin^2 \theta \partial^2_\theta G_\theta + \partial^2_\phi G_\theta +G_\theta - \sin\theta \cos \theta \partial_\theta G_\theta = 0
\ee
Proceeding like the previous case:
\bea
\partial_\theta c^m_\phi -2\cot\theta c^m_\phi&=&-im c^m_\theta
\label{thetaphi1}\\
\sin^2 \theta \partial_\theta c^m_\theta-\sin \theta \cos\theta c^m_\theta &=&im c^m_\phi
\label{thetaphi2} \\
\sin^2 \theta \partial^2_\theta c^m_\theta -m^2c^m_\theta +c^m_\theta- \sin\theta \cos \theta \partial_\theta c^m_\theta &=& 0
\label{thetaphi12}\\
\partial^2_\theta c^m_\theta &=& a c^m_\theta 
\label{thetaphi3}\\
im \partial_\theta c^m_\theta &=& b c^m_\phi
\label{thetaphi4}
\eea
If $m\ne 0$, from the previous equations:
\bea
\left(a \sin^2 \theta-m^2 +1\right)c^m_\theta+i \frac{b}{m}\sin\theta \cos \theta  c^m_\phi&=& 0\\
\left( \frac{b}{i m}\sin^2 \theta -\sin \theta \cos\theta \right) c^m_\theta -im c^m_\phi&=&0
\eea
The result is $c^m_\theta = c^m_\phi =0$ for all $m \ne 0$.\\
If $m=0$ and $a=-1$, we get $c^0_\theta= \sin \theta$ and $c^0_\phi=0$ and so $G_\theta= \sin \theta$ and $G_\phi=0$.\\
So we are left with the following equations:
\bea
\left[ \partial_\chi + \frac{\overline X_\rho}{ \overline X_\chi} \tan \rho - \frac{\overline X_\chi}{\overline X_\rho} \left( \partial_\rho- \tan \rho \right) \right] A_\rho  +\left[ \partial_\rho -2\tan \rho - \frac{\overline X_\rho}{ \overline X_\chi} \partial_\chi\right] A_\chi &=& 0 
\label{third1}\\
A_\rho-\frac{\overline X_\rho}{\overline X_\chi} A_\chi - \left[ \partial_\rho 
-2\tan\rho  -\frac{\overline X_\rho}{\overline X_\chi} \left( \partial_\chi 
-2\cot\chi \right) \right] A_\theta &=&0 
\label{third2}\\
\sin^{2}\chi \tan\rho A_\rho-\sin \chi \cos\chi A_\chi - A_\theta &=& 0
\label{third3}
\eea
Taking a linear combination of the second and the third equation we get:
$$
\left[ \partial_\rho 
-2\tan\rho  -\cot \rho \cot \chi \partial_\chi 
+2\cot^2\chi \cot \rho  -\cot\rho \sin^{-2} \chi \right] A_\theta =0 
$$
The solution to the previous system gives:
$$
\delta W_\theta = \sin \theta \sec \rho \tan \rho \sin \chi \;F\left[\frac{\cos \chi}{\sin\rho}\right]
$$
where $F(x)$ are generic functions.\\
Using the other two equations we get $F=0$: so in this case we get no solution since we have assumed that $\delta W_\theta \ne 0$.\\

To summarize the results for the static case,
the first and third possibility do not yield
any solution, while the second gives an infinite
family of solutions.
These solutions are generically regular at the horizon,
and therefore they cannot remove the singularity
of the background solution $\overline X_\mu$ there.
We conclude that all the solutions
of the system (\ref{con00}-\ref{con33})
in the neighborhood of (\ref{xstat})
are singular at the horizon.

\subsection{Flat slicing case}

\textit{First possibility:  $\delta c \ne 0 $}
\smallskip

The same reasoning that was used for the static case
leads to equations (\ref{deltac1},\ref{deltac2}).
Inserting (\ref{deltac2}) in (\ref{deltac1}) we obtain a complex solution, which is not acceptable.
\medskip

\textit{Second possibility: $\delta c = 0 $ and $\partial_\theta \delta W_\theta = 0$ }
\smallskip

The same reasoning that was used for the static case
leads to $\delta W_\theta=0$ and $\delta W_\phi=A_\phi\sin^2\theta$.

Also the analysis of equation (\ref{lin1}) proceeds in the same
way, leading to $G_\rho=G_\chi$.

Equation (\ref{lin2}) implies that $\partial_\theta G_\rho=0$.
Equation (\ref{lin3}) implies that $\partial_\phi G_\rho$
is a function of $\theta$, $\rho$ and $\chi$.
Thus $G_\rho$ must be linear in $\phi$,
but the only such function that is compatible with
periodicity is independent of $\phi$.
Thus $G_\rho$ is a constant that we can set to one
without loss of generality. 

Proceeding in the same way of the static case,
from equations (\ref{lin1},\ref{lin4}) we obtain again
equations (\ref{secpos1}-\ref{secpos2}). 
Substituting the explicit form of $\overline X_\mu$,
the solution to these equations is:
\bea
\delta W_\rho &=& \sec \rho \cos \chi \:F[\tan \rho -\sec \rho \cos \chi]\\
\delta W_\chi &=& \sec \rho \tan \rho \sin\chi \;F[\tan \rho -\sec \rho \cos \chi]
\eea
So the $\delta X$'s are:
\bea
\delta X_\rho &=& \frac{ \sin \rho - \cos \chi } {H}\cos \chi \sec^4 \rho\;F'[\tan \rho -\sec \rho \cos \chi] \\
\delta X_\chi &=& \frac{ \sin \rho - \cos \chi }{H } \tan \rho \sec^2 \rho \sin \chi \;F'[\tan \rho -\sec \rho \cos \chi]
\eea
Imposing the normalization condition (\ref{norcon})
we get that $F$ is a constant.

So the $\delta X$'s are zero: this is a consequence to the fact that $\delta W =(\sec \rho \cos \chi,\sec^2 \rho \sin \rho \sin\chi,0,0)$ is a Killing vector.
\bigskip

\textit{Third possibility: $\delta c = 0 $ and $\partial_\theta \delta W_\theta \ne 0$ }
\smallskip

The analysis of this case proceeds as in the static case
down to equations (\ref{third1}-\ref{third2}-\ref{third3}).
Solving (\ref{third3}) for $A_\theta$ and inserting in the other two,
and using the explicit form of $\overline X_\mu$,
we get the following equations:
\begin{equation*}
\begin{split}
\left[ \partial_\chi - \frac{ \cos \rho \sin \chi}{1-\cos \chi \sin \rho}  \partial_\rho + \left( \frac{1-\cos \chi \sin \rho}{ \cos \rho \sin \chi} + \frac{ \cos \rho \sin \chi}{1-\cos \chi \sin \rho} \right) \tan \rho  \right] A_\rho  +
\qquad\qquad\qquad\\ 
+\left[ \partial_\rho - \frac{1-\cos \chi \sin \rho}{ \cos \rho \sin \chi}  \partial_\chi -2\tan \rho \right] A_\chi = 0 
\qquad\qquad&
\\
\left[ \sin^2\chi \tan \rho\, \partial_\rho - \frac{1-\cos \chi \sin \rho}{ \cos \rho} \sin \chi \tan \rho \, \partial_\chi -  \left( \cos ^2 \chi  +2 \sin^2 \chi \tan^2 \rho + \tan \rho \sec \rho \cos \chi \right)\right] A_\rho+\\
\left[ -\sin \chi \cos \chi \,\partial_\rho + \frac{1-\cos \chi \sin \rho}{ \cos \rho} \cos \chi\, \partial_\chi + 2 \tan \rho \cos \chi \sin \chi \right] A_\chi=0
\qquad\qquad&
\end{split}
\end{equation*}

If we multiply the first equation by $\sin \chi \cos \chi$ and we sum the two equations, we get an equation where only $A_\rho$ appears, and that equation has the following solution:
$$
A_\rho = \sec \rho \sin \chi \;F[\sin \rho \cos \chi-\tan \rho]
$$
Then multiplying the first equation by $\sin \chi \tan \rho \sec \rho (1-\cos \chi \sin \rho)$ and summing the two equations, we get an equation where $A_\rho$ appears only algebraically. Substituting $A_\rho$ with the solution found above, we get the following solutions:
$$
A_\chi = \frac{2\cos \chi \sin \rho -1\pm i}{2\sqrt{2}\cos^2 \rho} \;F[\sin \rho \cos \chi-\tan \rho]
$$
We can note that it is not necessary to impose the normalization condition in order to conclude that the solution does not exist: in fact we have obtained a complex solution, so there is no acceptable solution in this case.

So the overall conclusion is that the solution (\ref{xflat}) is isolated.

\end{appendix}


\end{document}